\documentclass[%
 aip,
 amsmath,amssymb,
 reprint,%
]{revtex4-2}

\usepackage{graphicx}
\usepackage{dcolumn}
\usepackage{bm}

\usepackage[utf8]{inputenc}
\usepackage[T1]{fontenc}
\usepackage{mathptmx}
\usepackage{etoolbox}

\graphicspath{ {./figure/} }
\usepackage[colorlinks=true, linkcolor=blue, urlcolor=blue, citecolor=blue]{hyperref}
\usepackage{amsmath}  
\usepackage{amsfonts} %
\usepackage{amssymb}  %
\usepackage{subcaption}
\makeatletter
\def\@email#1#2{%
 \endgroup
 \patchcmd{\titleblock@produce}
  {\frontmatter@RRAPformat}
  {\frontmatter@RRAPformat{\produce@RRAP{*#1\href{mailto:#2}{#2}}}\frontmatter@RRAPformat}
  {}{}
}%
\makeatother
\begin{document}

\preprint{AIP/123-QED}

\title[]{Aerodynamics and Aeroacoustics of da Vinci's Aerial Screw}
\author{Suryansh Prakhar}
\author{Jung-Hee Seo}%

\author{Rajat Mittal}
 \email{mittal@jhu.edu.}
\affiliation{ 
\textsuperscript{1}Department of Mechanical Engineering, Johns Hopkins University, Baltimore, MD, USA
}%

\date{\today}

\begin{abstract}
Leonardo da Vinci’s aerial screw, conceived in the 15th century, represents one of the earliest conceptualizations of lift-generating rotary flight. Despite its historical significance, the aerodynamic and aeroacoustic performance of this rotor has received limited scientific attention. In this study, we employ direct numerical simulations to analyze the aerodynamic forces and acoustic emissions of a modernized da Vinci aerial screw design across a range of Reynolds numbers (2000, 4000, 8000, and 16000). These results are compared against those from a canonical two-bladed rotor producing similar lift. The aerial screw demonstrates 42.2\% lower mechanical power consumption and 72.3\% lower acoustic intensity per unit lift, primarily due to its larger wetted area and correspondingly lower rotational speed. Although the aerial screw exhibits a lower lift coefficient and much of its surface contributes minimally to lift generation, the net performance under iso-lift conditions highlights its efficiency and reduced noise signature. The continuous spiral geometry of the aerial screw also helps suppress blade–vortex interaction noise common in multi-bladed systems. These findings support previous scaling analyses and point toward unconventional rotor designs as viable options for low-noise aerial platforms. 
\end{abstract}

\maketitle

\section{Introduction}
Leonardo da Vinci, born in Anchiano, Italy in 1452, was a polymath whose contributions extended well beyond his renowned Renaissance paintings and sculptures. He produced detailed anatomical drawings, performed dissections, and made significant advances in geology, optics, hydrodynamics, and several other fields \citep{wikipediaLeonardoInventions}. Many of da Vinci’s engineering concepts remained unrealized during his lifetime, including his design for an “aerial screw,” envisioned as an early flying machine. A 3D reconstruction of this design is shown in Figure \ref{fig:daVinciGeometry}, featuring a large spiral membrane supported by a central shaft, struts, and tensioned cables. The base disk was intended to be operated by humans using handles attached to the shaft, generating rotational motion to lift the device into the air \cite{laurenza2006leonardo}.

While the human powered aspect of the design is impractical, the rotary motion of the aerial screw should and does generate lift \citep{acsir2020solidityone,elico}, thereby validating the overall concept behind da Vinci's design.
Indeed in addition to the studies of the aerodynamics of the aerial screw ~\citep{acsir2020solidityone,elico} da Vinci's
idea has also inspired several patents that use propellers designs similar to the aerial screw but these are primarily used for ships. For instance, \citet{US941923A} used a spiral propeller containing multiple turns with smaller radius at the ends of the propeller and the radius increases towards the center. \citet{US505402A} and \citet{US885109A} made use of the space at the other end of the spiral by fitting another spiral, thus using a double bladed screw propeller.

Beyond the aerodynamic performance of rotors/propellers, the rapid expansion of drones for a wide range of applications\cite{DroneNoiseRev} has also brought to fore the importance of the aeroacoustic noise generated by rotors. There has been a concerted effort in recent times to examine the source of aeroacoustic noise from these rotors and to create new designs with a lower acoustic signatures. 
\citet{yu_bvi} has listed several design modifications incorporated by traditional propellers to reduce the aeroacoustic noise. Some of these include modifying the propeller tip as in Ogee tip to prevent formation of a concentrated vortex, vane tips that have a notch at the trailing edge to produce tip vortex that are spaced apart and thus prevent blade vortex interactions, and using porous leading edge to reduce blade vortex interaction induced surface pressure fluctuation at the leading edge.
Apart from conventional propeller designs, \citet{cmyers} filled a patent for an improved ``screw propeller'' design in the 1890s for use in steamships. This propeller featured blades that formed a continuous loop near the tip. There are several propeller patents since then that have used looped end blades and more recently, propellers with looped end blades  have reported to produce better fuel efficiency and reduced noise for marine applications~\cite{sharrow} and better aeroacoustic performance in drones~\cite{toroid}.

 We recently explored the simple strategy of increasing the wetted area of rotor blades for reducing aeroacoustic noise~\citep{bioinspired_bnb}that was inspired by insect wings\cite{seo_mos_scale,Seo_mosq} and some historical work on propeller noise reduction\cite{vogeley1949}. The larger wetted area allows the rotor to generate a given amount of lift at a lower rotational speed (RPM) and since aeroacoustic noise intensity scales with the second power of the rotation rate, this has been found to reduce the overall intensity of aeroacoustic noise generated for a fixed amount of lift. Interestingly, the reduction in aeroacoustic noise intensity is accompanied with a reduction in the aerodynamic power requirement, thereby providing an additional benefit for these large area rotors.

Based on these results, we hypothesize that da Vinci's aerial screw, which has a very large wetted area compared to conventional rotors may also exhibit similar benefits in mechanical power and acoustic intensity reduction. The multi-blade design of traditional rotors can generate blade vortex interaction noise from the interaction of vortices shed from one blade with the other blades\citep{brentner1994helicopter}. The single continuous blade of da Vinci's aerial screw may be able to diminish this noise mechanism as well. The objective of the current study is to employ direct numerical simulations of the flow and aeroacoustics to examine our hypothesis. We consider the effectiveness of the aerial screw quantified in terms of acoustic noise characteristics and aerodynamic power consumption and the acoustic noise, and compare these to canonical drone rotors.

\section{Methodology}
\subsection{Rotor Geometry} \label{ssec:rotorGeo}
\begin{figure*}
\centering
\includegraphics[width=0.8\textwidth]{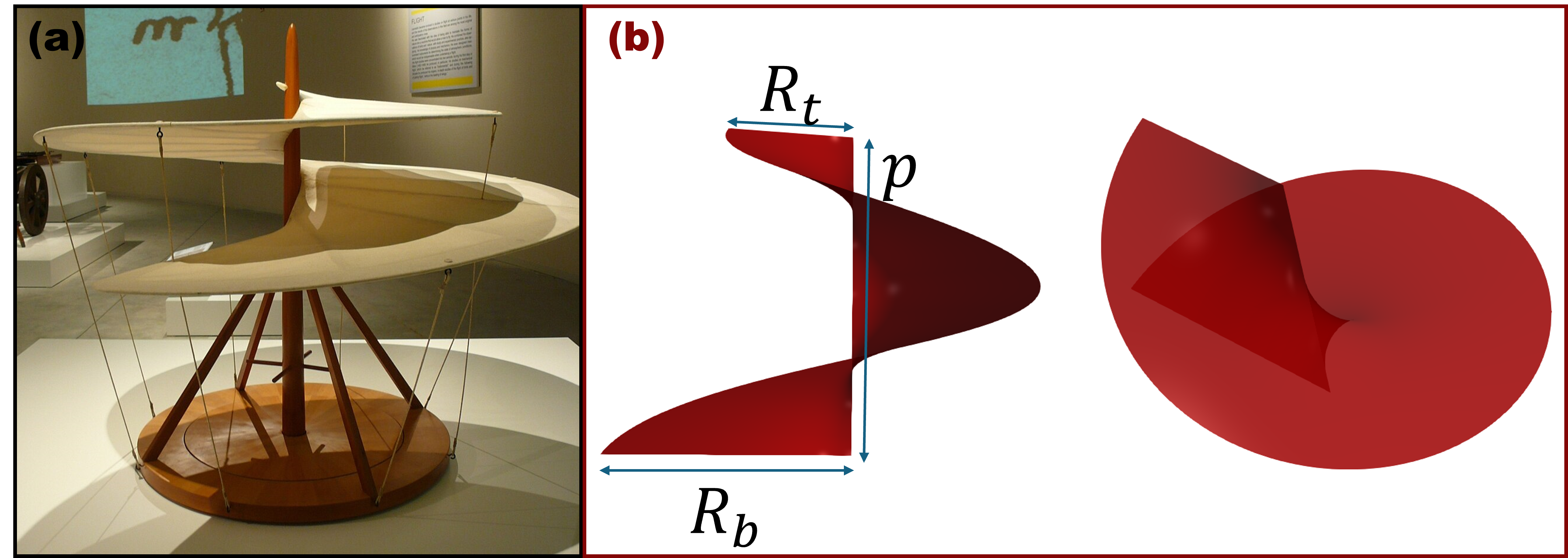}
\caption{\centering (a) A 3D model of the da Vinci's aerial screw taken from Wikipedia~\citep{wiki_permission_aerial_screw}. (b) The CAD model of the da Vinci's aerial screw having a taper ($R_t/R_b$) of 1:2, $p/R_b=1.31$ and one full turn used in the current study. This model is based on the parameters used by \citet{elico}.}
\label{fig:daVinciGeometry}
\end{figure*}

A 3D model of the da Vinci's aerial screw sketch is shown in figure \ref{fig:daVinciGeometry}a and we observe a spiral rotor blade with a large base radius that tapers as it ascends. The spiral has one and a half turn and due to the design of the rotor, the angle of attack varies across different sections of the rotor.
There is a limited research on the aerodynamics of the da Vinci's rotor with one of the parametric study from \citet{elico} that explored various parameters such as the rotor radius, pitch, number of turns, taper, lip and the anhedral angle. The authors of the study noted that the rotor performance depends on the pitch to radius ratio and they selected their final geometry to have a pitch to base radius ratio of 1.31. The top to base radius ratio (taper) was set to 1:2 as the study found that the tapered rotors were more efficient at producing lift. Furthermore, the total number of turns was set to 1 which differs from the design shown in figure \ref{fig:daVinciGeometry}a. The study shows that increasing the number of turns for tapered rotor can increase the lift due to increased area but at the same time, the mechanical power requirement is also increased. The da Vinci's aerial screw design used in this paper is based on these parameters and the CAD model of the rotor geometry is shown in figure \ref{fig:daVinciGeometry}b. 

\subsection{Flow Solver}
We use our in-house incompressible Navier-Stokes solver called ViCar3D that has been used in several previous studies such as \citet{mittal2008,seo2011,new_ibm,Seo_mosq,ji_fish_1,ji_fish_jfm,aghaei2022contributions}. We couple the continuity equation,
 \begin{equation}
     \nabla \cdot {\bf u} = 0 \,\,\, ,
     \label{conteqn}
 \end{equation}
 and the momentum equation in non-inertial rotating reference frame~\citep{batchelor_book,speziale_rrf_eqn},
 \begin{equation}
     \frac{\partial {\bf u}}{\partial t} + {\bf u}\cdot \nabla {\bf u}  = -\frac{1}{\rho}\nabla P +\nu \nabla^2 {\bf u} -2\bm{\Omega} \times {\bf u} - \bm{\Omega} \times (\bm{\Omega} \times {\bf x}) \,\,\,,
\label{momeqn}
 \end{equation}
 to obtain the velocity and the pressure field. The last two terms of equation \ref{momeqn} account for the forces acting on the non-inertial frame with $2{\bm{\Omega}}\times {\bf u}$ being the Coriolis acceleration term and $\bm{\Omega} \times (\bm{\Omega} \times {\bf x})$ corresponding to the centrifugal acceleration term (here, $\bm{\Omega} = \left(  0, 0, \Omega \right) $ is the rotation vector of the reference frame).

 The body is constructed using a zero thickness membrane with a surface described by an unstructured grid made of triangular elements while the flow simulations is performed on a non-uniform grid Cartesian grid shown in fig. \ref{fig:fine_mesh}. The momentum equation is decomposed into an advection-diffusion equation and a pressure Poisson equation using Van-Kan's fractional step method~\citep{vankan}. The Poisson equation is solved using a biconjugate stabilized gradient descent method and for the advection-diffusion equation, a second order Adam-Bashforth scheme is used for the advection term and an implicit Crank-Nicolson scheme used for viscous term.

 The pressure lift is then computed using,
 \begin{equation}
      F_L= -\int_S P n_z dS \,\,\, ,
 \end{equation}
where $S$ is the blade surface and $n_z$ is the surface normal pointing in the lift direction.
The lift coefficient is defined as,
 \begin{equation}
      C_L= \frac{F_L}{(1/2) \rho v_t^2 A} \,\,\, ,
 \end{equation}
 where $v_t$ is the blade tip velocity and $A$ is the total blade area. The mechanical power required for the blade to work against the pressure loading is calculated as,
\begin{equation}
     W_M = \int_S P \left( {\bf n} \cdot {\bf v} \right) dS \,\,\, ,
\end{equation}
where, surface velocity is denoted using ${\bf v}$ and ${\bf n}$ is the surface normal.
 
\subsection{Aeroacoustics Solver}
We use the Brentner and Farassat integral formulation of the Ffowcs Williams-Hawkings equation~\citep{FWHorigpaper,fwheqn2,brentner_fwh_term_def} for aeroacoustic noise given by:
\begin{equation}
\begin{split}
        4\pi p^{\prime}({\bf x},t) = & \frac{1}{c}\int \Big[\frac{\dot{L}_r}{r(1-M_r)^2} \Big]_{t-r/c} dS \\
    &+\int \Big[\frac{L_r-L_M}{r^2(1-M_r)^2} \Big]_{t-r/c} dS \\
    & +\frac{1}{c}\int \Big[\frac{L_r(r\dot{M_r}+c(M_r-M^2))}{r^2(1-M_r)^3} \Big]_{t-r/c} dS \,\, .
    \end{split}
    \label{fwheqn}
\end{equation}
Here, the sound pressure is denoted using $p'$, the recording location is placed at a distance of $r$ and the speed of sound is denoted using $c$. The Mach number, $ {\bf M}$, is defined as ${\bf v}/c$ with ${\bf v}$ being the surface velocity. $M_r$ denotes the Mach number vector in the direction of the monitoring point and is calculated as $ {\bf M}\cdot r$. The force vector is denoted using $ {\bf L}$ and computed using $P{\bf n}$ with $P$ and ${\bf n}$ denoting the surface pressure and normal. $L_r$ is defined as ${\bf L}\cdot {\bf r}$ and  $L_M={\bf L}\cdot {\bf M}$.

Once we have the sound pressure, the directivity is obtained using the RMS value of the sound pressure ($p'_\text{rms}$)
and the acoustic intensity is calculated using, 
\begin{equation}
    I_a=  \frac{1}{4 \pi r^2 \rho c} \int_S (p_{rms}^{'})^2 dS \,\,\, .
    \label{AcousticInt}
\end{equation}

\subsection{Scaling Analysis}
\label{sec:scaling}

The da Vinci's aerial screw will be compared against a canonical two bladed rotor to compare the mechanical power and acoustics when both these rotors produce similar lift. We know from our previous scaling analysis~\cite{bioinspired_bnb} of rotor with flat blades with a pitch angle of $\theta$ that the lift force can be expressed as, 
\begin{equation}
    F_L  = \frac{1}{2} \rho R_b^2 \Omega^2 A  C_L \,\, ,
    \label{eqn:forcescale}
\end{equation}
where, $R_b$ is the rotor radius. Since the net lift being produced is fixed, the rotation speed will scale with blade area as,
\begin{equation}
 \Omega^2 \sim \frac{1}{A}  \frac{1}{C_L}\,\, .
 \label{eqn:omega_area}
\end{equation} 
The drag force can be written as,
\begin{equation}
    F_D =F_L \tan{\theta} \,\, ,
    \label{lifteqn}
\end{equation}
and the mechanical power can be expressed as,
\begin{equation}
    W_M \sim F_D R_b \Omega  \sim R_b^3 \Omega^3 A  C_L    \tan{\theta} \,\, ,
    \label{eqn:power_omega}
\end{equation}
where $F_DR_b$ represents the aerodynamic torque. Therefore, using equation \ref{eqn:omega_area}, the relation between the aerodynamic power and rotor area for a rotor with flat blades is,
\begin{equation}
    W_M \sim A^{-\frac{1}{2}} C_L^{-\frac{1}{2}}  \tan{\theta}\,\, .
    \label{eqn:power_scaling}
\end{equation}
The acoustic sound pressure ($p'$) will scale as~\citep{seo_mos_scale},
\begin{equation}
    p' \sim \frac{dF}{dt} \sim  R_b^2 \Omega^3 A  C_L    \tan{\theta}  \,\,\, ,
\end{equation}
and since $I_a \sim p^{'2}$, using equation~\ref{eqn:omega_area}, we can write the scaling of acoustic intensity with area as, 
\begin{equation}
    I_a \sim \frac{1}{A}  \frac{1}{C_L}\,\,\, .
    \label{eqn:intensity_scaling}
\end{equation}
Thus, both the aerodynamic power and acoustic intensity reduces with increasing rotor wetted area, as long as the $C_L$ is not adversely affected by the difference in rotor shape. Indeed, we expect that due to the very different geometry of the aerial screw rotor, the lift coefficient would be quite different from that of a canonical rotor. Additionally, the area and the pitch angle, as well as the chord changes along the spiral of the aerial screw, so the above scaling laws that work for rotors with flat blades will not directly work for the aerial screw. Nevertheless, we hypothesize that despite the more complex rotor shape, the larger rotor area of the aerial screw compared to conventional rotor blades will still yield some reductions in aerodynamic power and acoustic intensity, and examining this hypothesis is the focus of our modeling study.

\section{Results}
\subsection{Cases}
As described in section \ref{ssec:rotorGeo}, the rotor is modeled with a base radius of $R_b$, top radius of $0.5R_b$, a pitch of $1.31R_b$ and contains 1 turn. This paper will compare four cases with Reynolds number of 2000, 4000, 8000, and 16000, where the Reynolds number is described based on the tip velocity ($v_t$) at the base of the rotor and the base radius ($R_b$). 
The angular velocity of rotation is set as $v_t/R_b$ for the case with Reynolds number of 8000, and the angular velocity halved, quartered, and doubled for the cases with Reynolds number 4000, 2000, and 16000 respectively. Note that much higher Reynolds number (such as 200,000 which would be typical for actual small size drones) are quite difficult to simulate with direct numerical simulations and we have therefore limited ourselves to comparing the designs at these lower Reynolds numbers. 

\begin{figure*}
\centering
\includegraphics[width=0.7\textwidth]{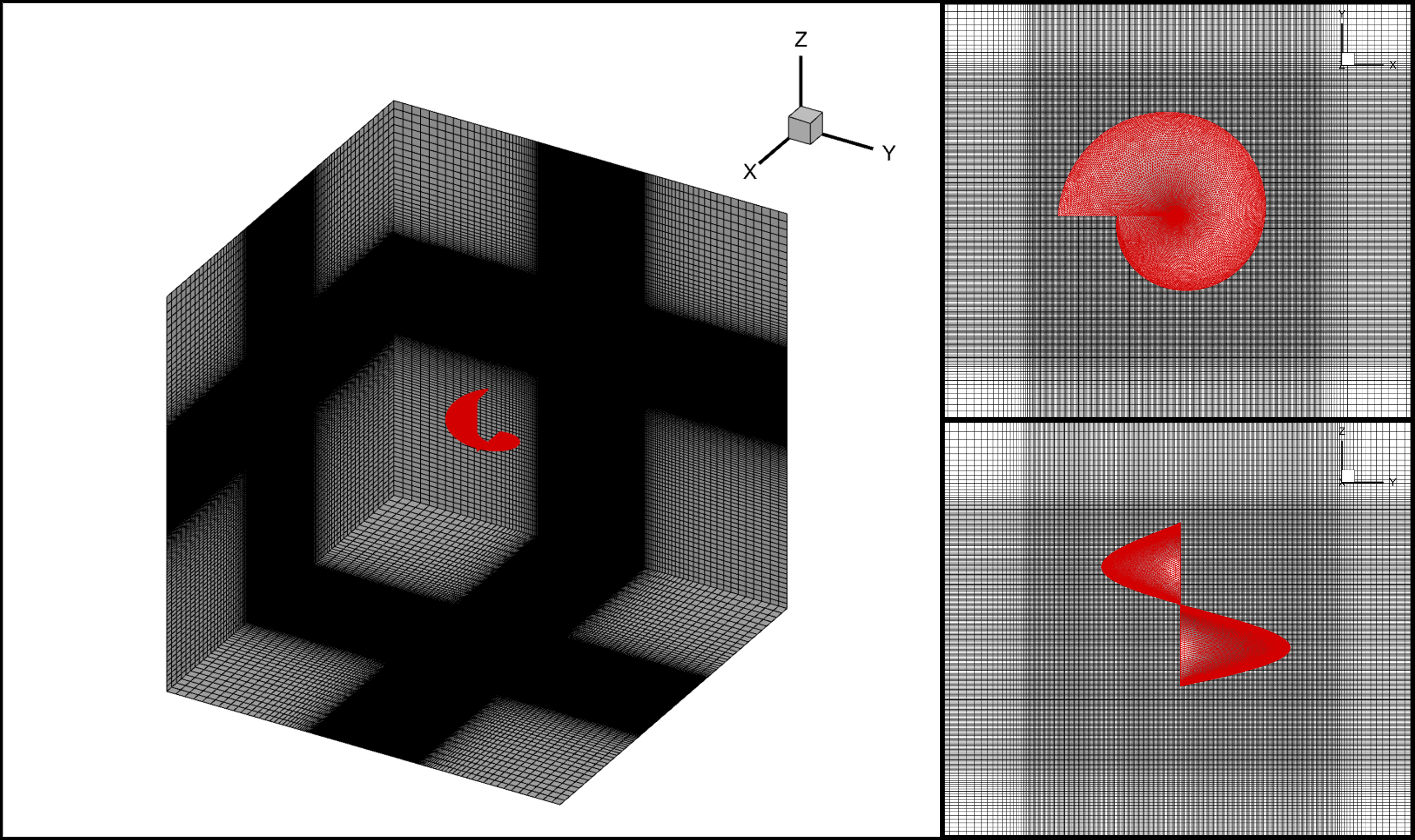}
\caption{\centering The fine mesh with 32 million grid points is shown in the flow domain along with the da Vinci rotor. A uniform fine mesh is used near the rotor body with a slowly expanding mesh away from the body.} 
\label{fig:fine_mesh}
\end{figure*}

Fig \ref{fig:fine_mesh} shows the topology of the grid used in these simulations. The region around the rotor is placed in a uniform fine mesh with a gradually expanding mesh used away from the rotor body. In the plane of the rotation, a $2.4R_b\times 2.4R_b$ region centered around the rotor is finely resolved while in the vertical axis, a length of $2R_b$ surrounding the screw is finely resolved. Based on the grid convergence study described in the appendix \ref{app:grid_convergence}, the Reynolds number 8000 and 16000 case employs a fine (32 million point) grid  while those with Reynolds number of 4000, and 2000 employ a medium (13.8 million point) grid.

The aim of this study is to also compare the effectiveness of aerial screw with canonical drone rotors, specifically when used to carry similar payload. We take a canonical two blade flat plate rotor from our previous work~\citep{bioinspired_bnb}, with the aspect ratio of the blades equal to 5, the angle of attack is set to 20$^\circ$ and the distance from rotation center to outer radius is denoted using $R_b$. Similar to the aerial screw, the rotor is modeled as a zero thickness membrane and is simulated in a rotating reference frame. We simulate three cases with angular velocities of $v_t/R_b$, $1.51v_t/R_b$, and $2v_t/R_b$ and these cases correspond to a tip velocity based Reynolds number of 8000, 12118 and 16000 respectively. The input parameters corresponding to all 7 cases are summarized in table \ref{table:summary_input_params}. These simulations were run for a total of four rotation cycles with last two cycles used to compute the mean lift and power. As will become evident in the next section, the reason for using a higher Reynolds numbers for the two bladed rotor is to obtain a case that produces similar absolute lift as that of aerial screw for a one-to-one comparison of the performance.

\begin{table}
\caption{\label{table:summary_input_params}The input Reynolds number and the rotation speed for all the seven cases.}
\begin{ruledtabular}
\begin{tabular}{lcc}
\textbf{Rotor} & \textbf{Re} &  \textbf{$\Omega$} \\
\hline
Aerial Screw   & 2000 & $0.25 v_t/R_b$ \\
Aerial Screw   & 4000 & $0.5 v_t/R_b$ \\
Aerial Screw   & 8000 & $v_t/R_b$\\
Aerial Screw   & 16000 & $2 v_t/R_b$\\
\hline
Canonical Rotor   & 8000 & $v_t/R_b$\\
Canonical Rotor    & 12118 & $1.51 v_t/R_b$\\
Canonical Rotor    & 16000 & $2 v_t/R_b$\\
\end{tabular}
\end{ruledtabular}
\end{table}

\subsection{Flow Features}
\begin{figure*}
\centering
\includegraphics[width=0.75\textwidth]{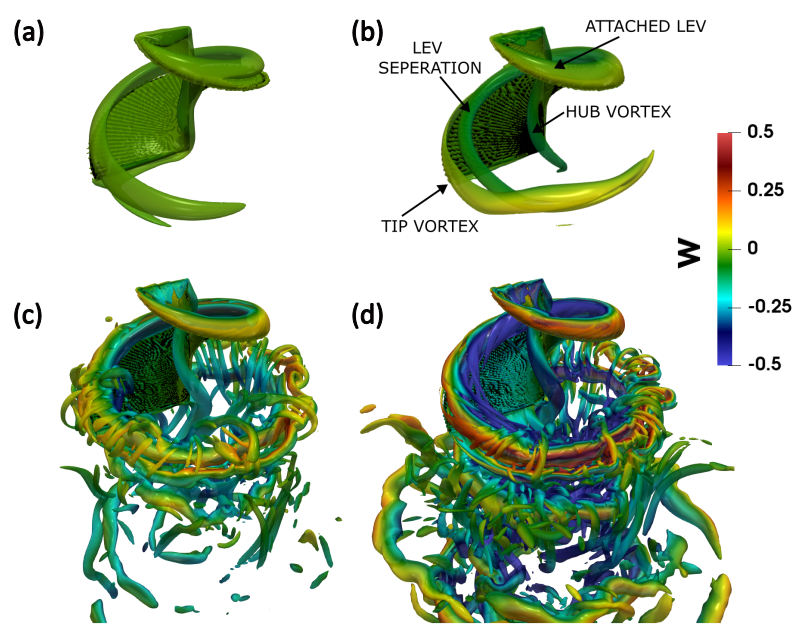}
\caption{\centering The vortex structures shown using iso-surface of $Q$ for Reynolds number (a) 2000, (b) 4000, (c) 8000, and (d) 16000 cases and colored using the vertical velocity component. These are at a late stage in the simulation corresponding to $t/T=5.9$} 
\label{fig:all_contour}
\end{figure*}

The four aerial screw cases with Reynolds number of 2000, 4000, 8000, and 16000 are simulated for a total of 6 cycles to make sure the cycle to cycle variation in the mean lift force has converged. The flow field is shown in figure \ref{fig:all_contour} with the vortices shown using isosurface of Q.
The case with Reynold number 2000 shows an attached leading edge vortex over almost the entire spiral leading edge. This vortex  extends only short distance into the the wake before dissipating. There is also a small tip vortex that extends into the near wake. The case with Reynolds number of 4000 also shows an attached LEV that separates noticeably before the end of the spiral leading edge. A tip vortex is formed at the end of the spiral leading edge and the two vortices spiral down in the downwash of the aerial screw for later times. This case also shows the presence of a hub vortex that spirals down the hub of the aerial screw. Finally for a Reynolds number of 8000, the LEV detaches from the leading-edge a bit earlier and similar to the Re=4000 case, and we see both the detached LEV and the tip vortices spiral into the downwash. However, at higher Reynolds numbers of 8000 and 16,000, the LEV experience an instability wherein smaller vortices are formed around the core of the large LEV. These instabilities amplify in the wake where there is a complete breakdown of the spiral vortex structures. This case also exhibits a hub vortex.

\begin{figure*}
    \centering
 \begin{subfigure}[b]{0.32\textwidth}
      \centering
         \includegraphics[width=\textwidth]{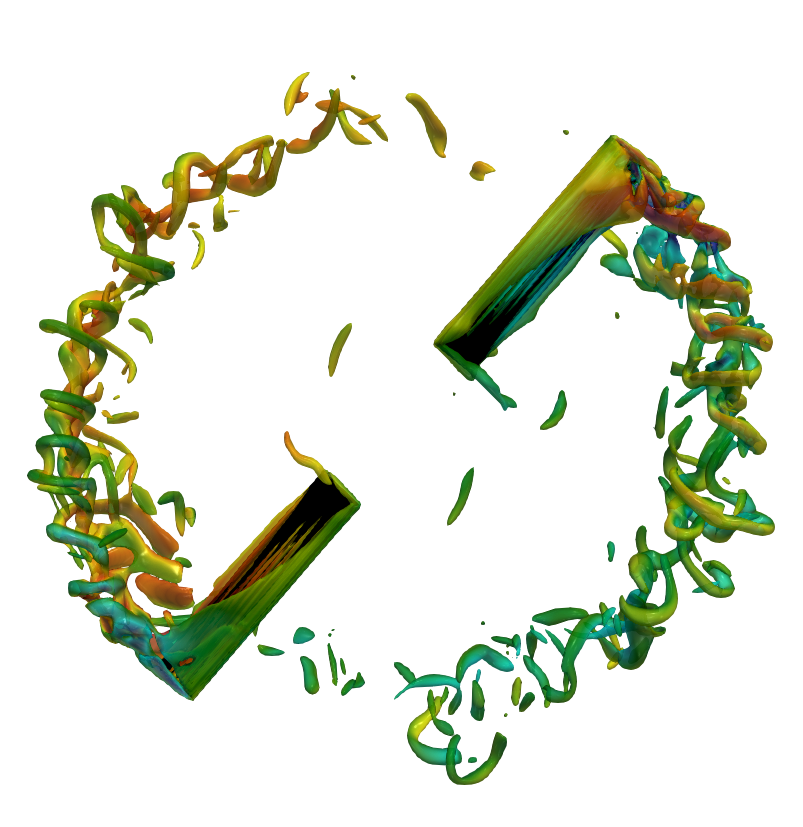}
         \caption{}
         \label{fig:CanonicalRotorQcontour:a}
     \end{subfigure}
  \hfill
     \begin{subfigure}[b]{0.32\textwidth}
         \centering
         \includegraphics[width=\textwidth]{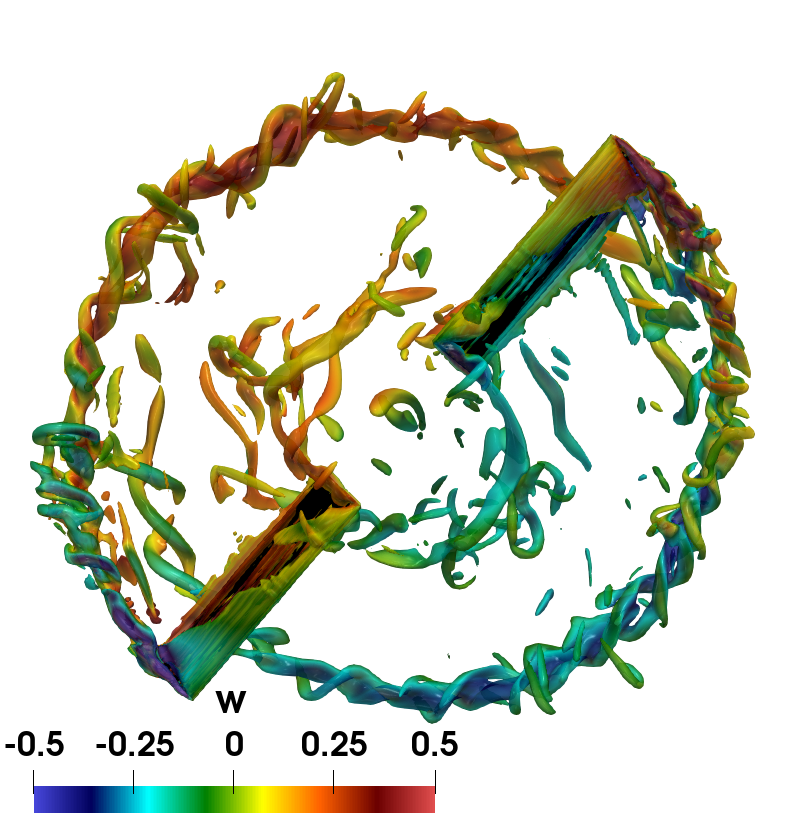}
         \caption{}
         \label{fig:CanonicalRotorQcontour:b}
     \end{subfigure}
      \hfill
     \begin{subfigure}[b]{0.32\textwidth}
         \centering
         \includegraphics[width=\textwidth]{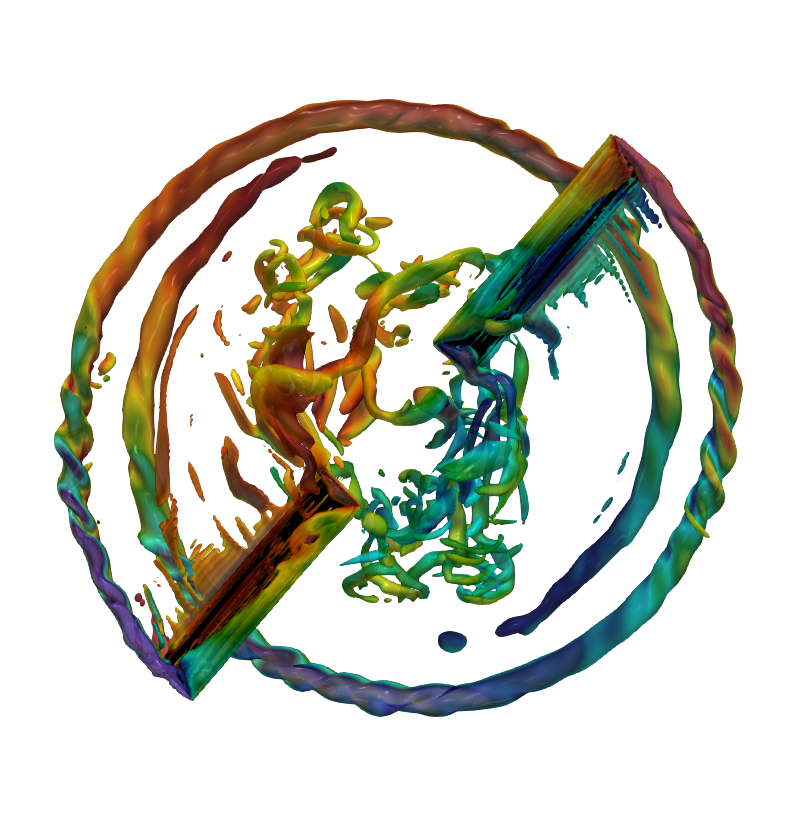}
         \caption{}
         \label{fig:CanonicalRotorQcontour:c}
     \end{subfigure}
\caption{\centering The flow field of the two blade rotors shown using iso-surface of the Q  and colored using the vertical velocity component at $t/T=4.0$, corresponding to Reynolds number of (a) 8000, (b) 12118 and (c) 16000.}
\label{fig:CanonicalRotorQcontour}
\end{figure*}

The vortex structures for the two-bladed conventional rotor are shown in \ref{fig:CanonicalRotorQcontour}.  For all cases, there is a leading-edge vortex that grows in strength toward the rotor tip. The wake of all the cases are dominated by the tip vortices that spiral down into the wake. All the tip vortices indicate the presence of multiple helical vortex codes that wind around each other and at the highest Reynolds number, the tip vortex is observed to extend far into the wake of the rotor.

\subsection{Aerodynamic Forces}
\begin{figure*}
     \begin{subfigure}[b]{0.24\textwidth}
         \centering
         \includegraphics[width=\textwidth]{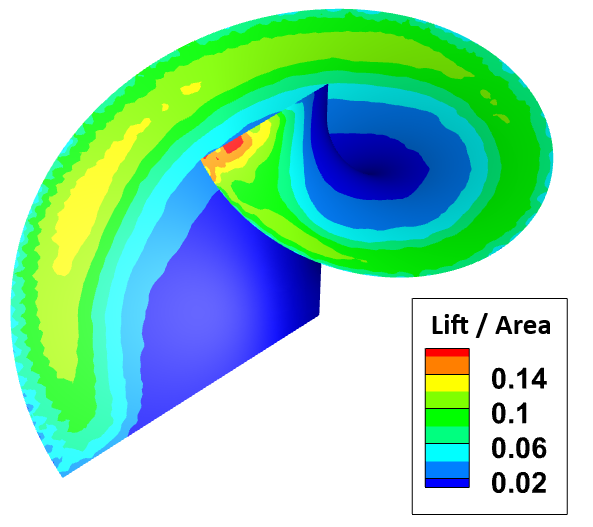}
         \caption{}
         \label{fig:avg_marker:a}
     \end{subfigure}
      \hfill
     \begin{subfigure}[b]{0.24\textwidth}
         \centering
         \includegraphics[width=\textwidth]{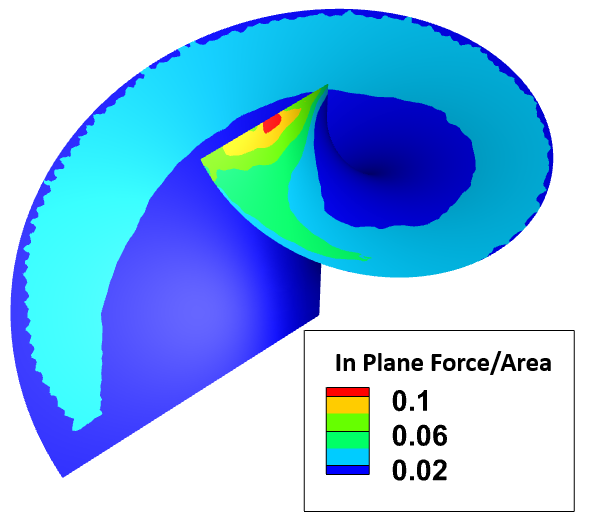}
         \caption{}
         \label{fig:avg_marker:b}
     \end{subfigure}
  \hfill
     \begin{subfigure}[b]{0.24\textwidth}
         \centering
         \includegraphics[width=\textwidth]{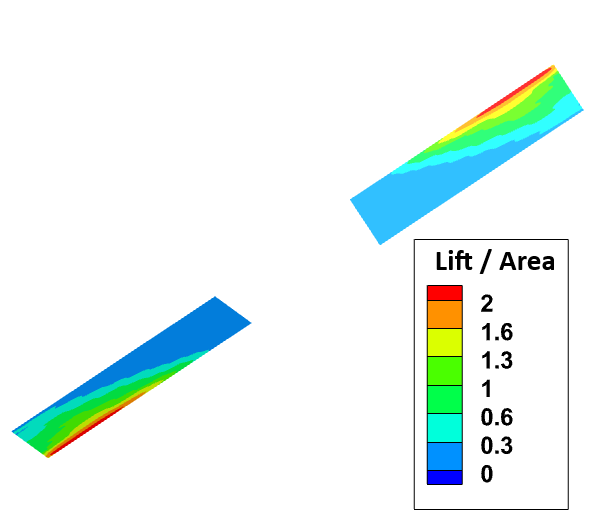}
         \caption{}
         \label{fig:avg_marker:d}
     \end{subfigure}
      \hfill
     \begin{subfigure}[b]{0.24\textwidth}
         \centering
         \includegraphics[width=\textwidth]{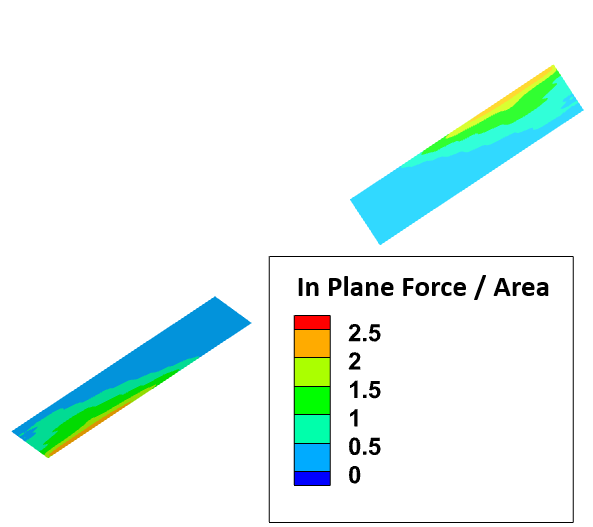}
         \caption{}
         \label{fig:avg_marker:e}
     \end{subfigure}
        \caption{\centering The averaged (a) lift density and (b) drag density contours for aerial screw at Reynolds number 8000 case and the averaged (c) lift density and (d) drag density  contours for canonical rotor at Reynolds number 12118 case.}
\label{fig:avg_marker}
\end{figure*}

We begin the discussion of the aerodynamic force by plotting in Figs. \ref{fig:avg_marker}a and b respectively, the average lift area-density (i.e. lift per unit area) and in-plane force area-density for the Re=8000 aerial screw over the surface of the aerial screw.  Additionally, Fig. \ref{fig:avg_marker}c and \ref{fig:avg_marker}d shows the same quantities for the canonical rotor at Reynolds number 12118 for comparison. For the canonical rotor, we note that the lift and drag force are primarily concentrated at the leading edge of the blade near the tip region. This is directly connected with the LEV over the canonical blade noted in the previous section. The plot for the aerial screw aerial screw indicates two regions of high lift area-density. One is at the top starting edge of the screw and this distribution is qualitatively similar to that for the conventional rectangular blade. The other region of high lift density is on the lower portion of the spiral near its leading-edge and this is associated with the leading-edge vortex on the spiral leading-edge noted in the previous section. The leading-edge velocity of the rotor is high at this location given the large radius of the screw and this enhances the strength of the LEV at this location. The in-plane force density shares these qualitative features as well. We note that large sections of the aerial screw rotor do not contribute to the generation of force.

\begin{figure*}
    \centering
 \begin{subfigure}[b]{0.45\textwidth}
      \centering
         \includegraphics[width=\textwidth]{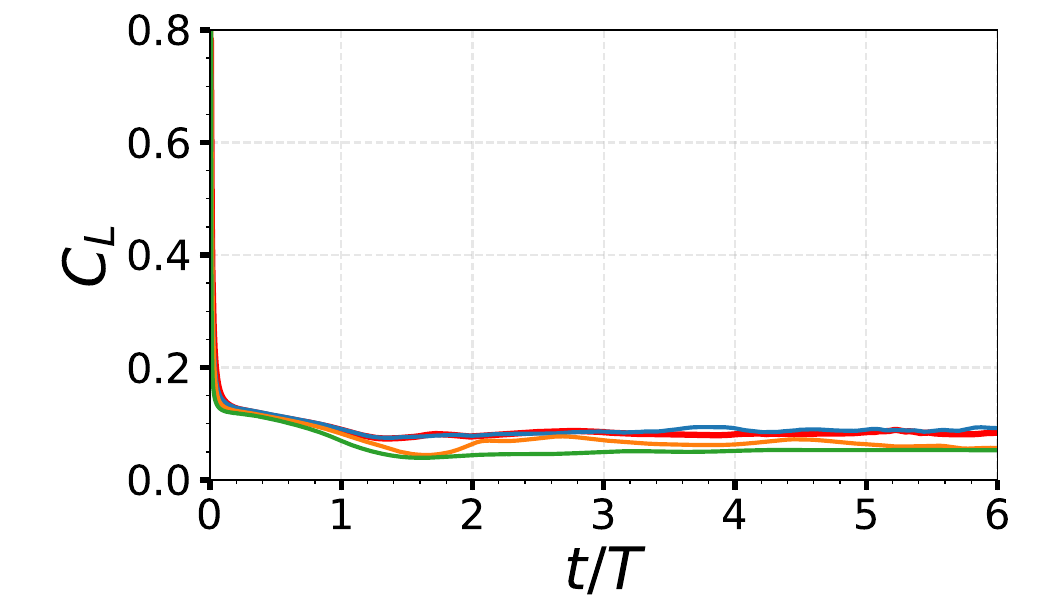}
         \caption{}
         \label{fig:lift_compare:a}
     \end{subfigure}
     \begin{subfigure}[b]{0.45\textwidth}
         \centering
         \includegraphics[width=\textwidth]{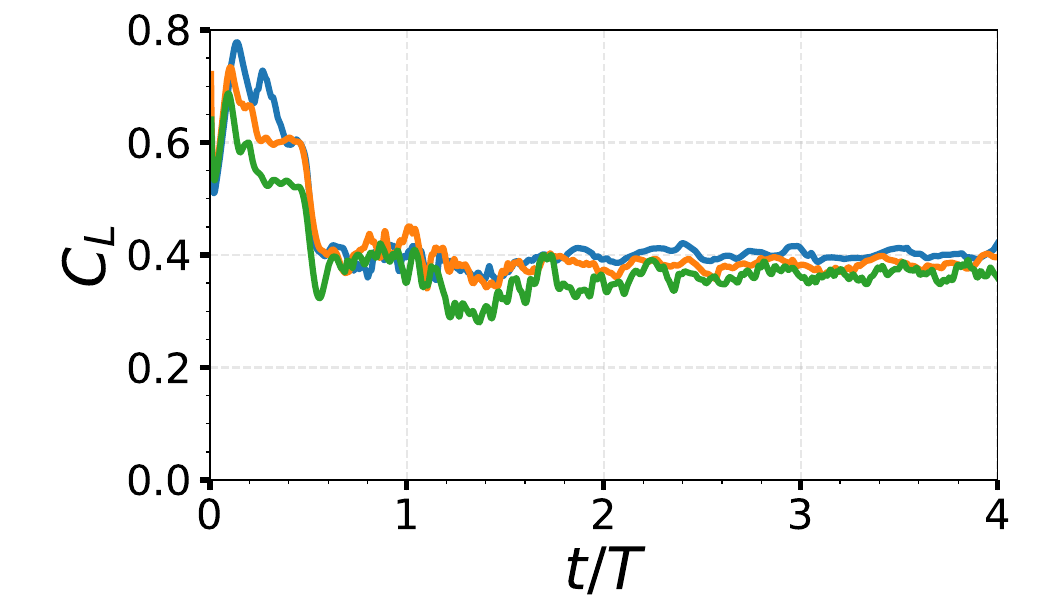}
         \caption{}
         \label{fig:lift_compare:b}
     \end{subfigure}
      \hfill
     \begin{subfigure}[b]{0.45\textwidth}
         \centering
         \includegraphics[width=\textwidth]{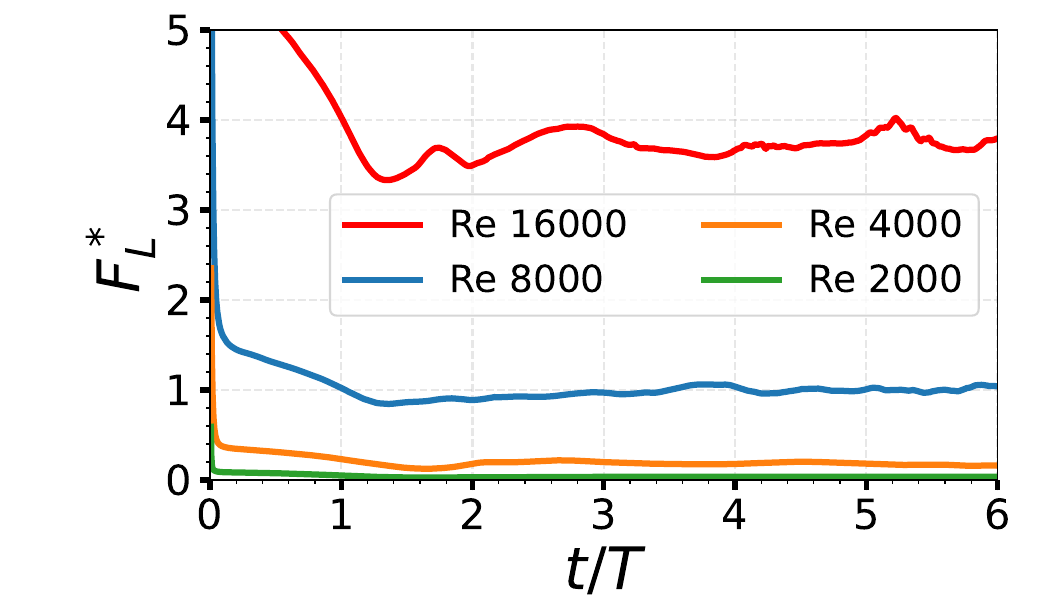}
         \caption{}
         \label{fig:lift_compare:c}
     \end{subfigure}
     \begin{subfigure}[b]{0.45\textwidth}
         \centering
         \includegraphics[width=\textwidth]{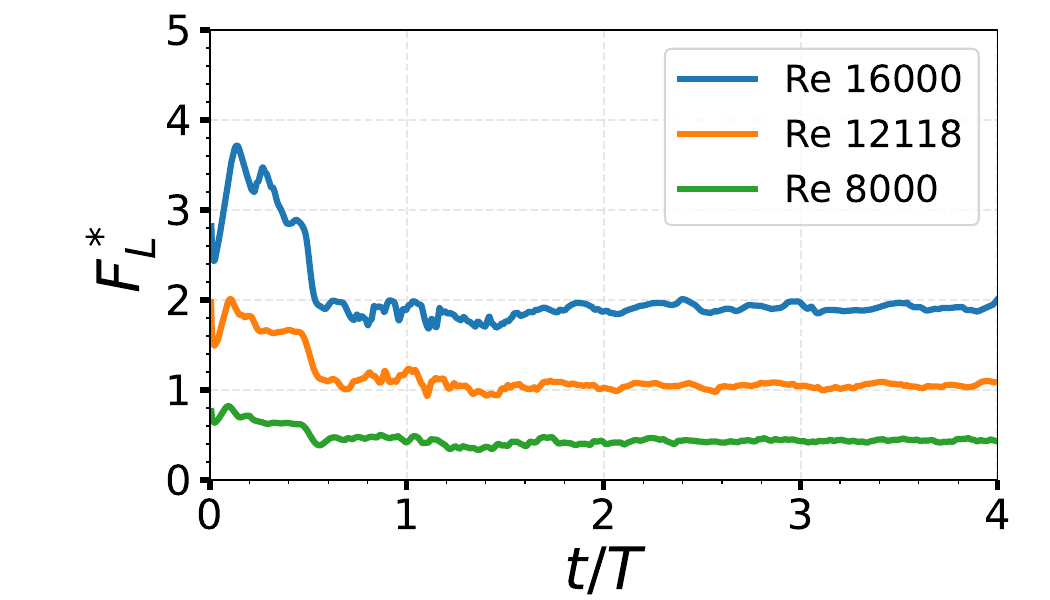}
         \caption{}
         \label{fig:lift_compare:d}
     \end{subfigure}
\caption{\centering The lift coefficient for the (a) aerial screw and the (b) the canonical rotor. Lift force produced by (c) aerial screw and (d) canonical rotor. The lift force is normalized using the mean lift of the Re=8000 case for the aerial screw.}
\label{fig:lift_compare}
\end{figure*}

The lift coefficient for the aerial screw (calculated based on the maximum tip velocity along the screw leading edge) is shown in figure \ref{fig:lift_compare:a} and that for the canonical rotor (based on its maximum velocity which is at the tip of the blade) is shown in figure \ref{fig:lift_compare:b}. First, we note that for these relatively high Reynolds number, the lift coefficient for both rotors are mostly independent of the Reynolds number. The lift coefficient for the canonical rotor is 137.5\% larger than that for the aerial screw, indicating that per unit wetted area of the rotor, the canonical rotor is more effective in generating lift than the aerial screw. This is due to the fact observed previously that the large portion of the aerial screw rotor does not contribute to the generation of lift, and furthermore, large portions of the blade are moving at a velocity significantly lower than maximum blade velocity. We note however, that the fact that the aerial screw \emph{does} generates positive lift for a wide range of rotation speeds validates the overall concept of the da Vinci aerial screw and his intuitive understanding of how the screw would impart downward momentum to the flow, and as a reaction, generate lift. 

While the lift coefficient is important in that it informs us about effectiveness of a rotor to generate lift, for practical considerations we are interested in comparing the power and aeroacoustic emission from rotors that are generating equivalent lift force (and therefore can lift the same payload).  The rotors are dimensionalized using the same  base radius ($R_b$). The lift forces for all the aerial screw and canonical rotor cases are shown in figure \ref{fig:lift_compare:c} and \ref{fig:lift_compare:d} respectively and the force values are normalized using the mean lift for the aerial screw with Reynolds number 8000 case. As one would expect, the higher Reynolds number cases with larger rotation speed generates higher lift with larger force fluctuations due to increased vortex shedding at higher Reynolds number. Among the seven cases simulated here the aerial screw at Re=8000  generated roughly the same amount of mean lift as the canonical rotor at Re= 12118 (the difference is about 4\%). Indeed, this match is not a coincidence since we first simulated the Re=8000 and 16000 cases for the canonical rotor and predicted the Reynolds number at which we would expect a lift equal to that of the aerial screw at Re=8000.

\begin{figure*}
    \centering
 \begin{subfigure}[b]{0.45\textwidth}
      \centering
\includegraphics[width=\textwidth]{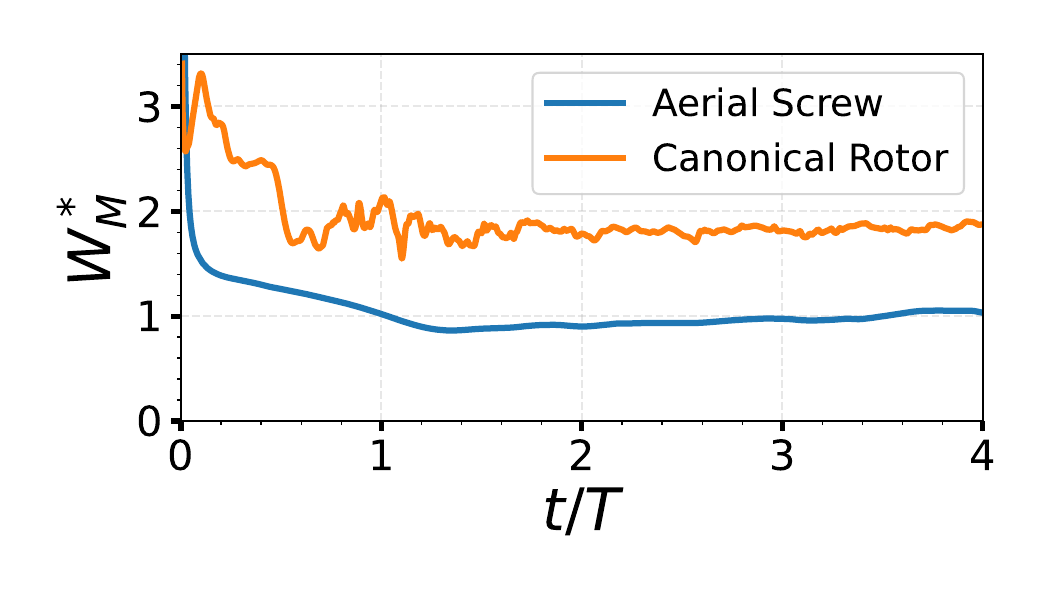}
         \caption{}
         \label{fig:power_compare:a}
     \end{subfigure}
     \begin{subfigure}[b]{0.45\textwidth}
         \centering
\includegraphics[width=\textwidth]{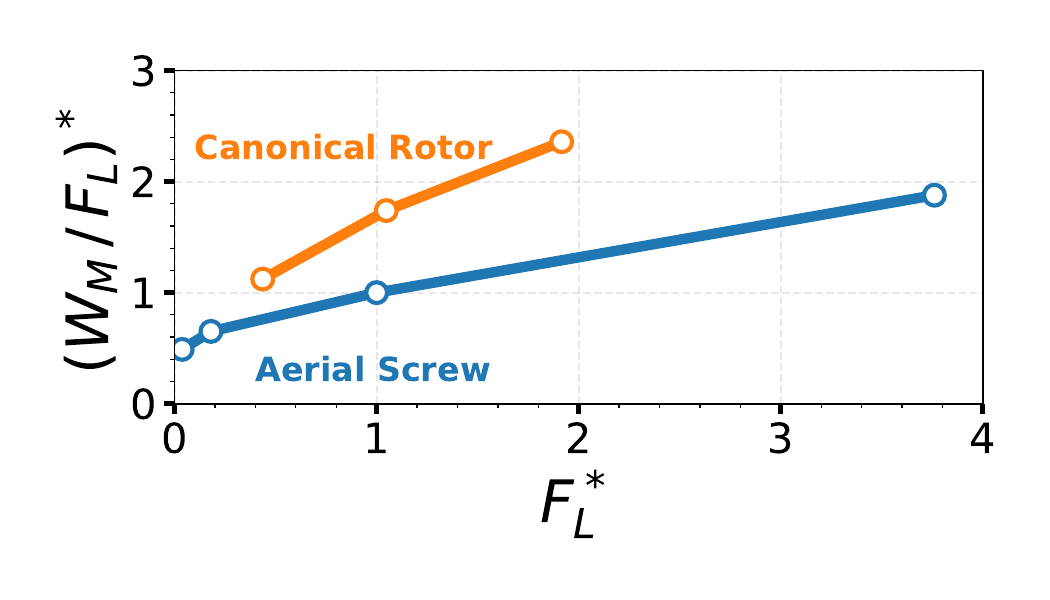}
         \caption{}
         \label{fig:power_compare:b}
     \end{subfigure}
\caption{\centering (a) The mechanical power is shown for for the two different rotors in iso-lift configuration. The aerial screw corresponds to Reynolds number of 8000 and the canonical rotor corresponds to Reynolds number of 12118 with the  mechanical power normalized using the mean power of the aerial screw with Re=8000 case. (b) Characteristic curve showing mechanical power vs mean lift for both the rotors and the values are normalized using the aerial screw with Re=8000 case.}
\label{fig:power_compare}
\end{figure*}

The mechanical power for both cases is shown in Figure \ref{fig:power_compare:a}, with power normalized by the mean power of the aerial screw. Although the aerial screw has a larger surface area—which could lead to increased drag—its lower rotational speed relative to the canonical rotor results in a net reduction in mechanical power consumption. This is consistent with the scaling relationship in Eq. \ref{eqn:power_omega}, where power varies with the cube of the rotational velocity. 

Figure \ref{fig:power_compare:b} presents the characteristic curve of mechanical power required for various lift magnitudes. The two rotors exhibit markedly different trends in specific power, suggesting that the aerial screw remains more power-efficient than the canonical rotor across a range of Reynolds numbers, including higher regimes.

\subsection{Aeroacoustics}
\label{sec:aeroacoustic}
\begin{figure*}
\centering
\includegraphics[width=1\textwidth]{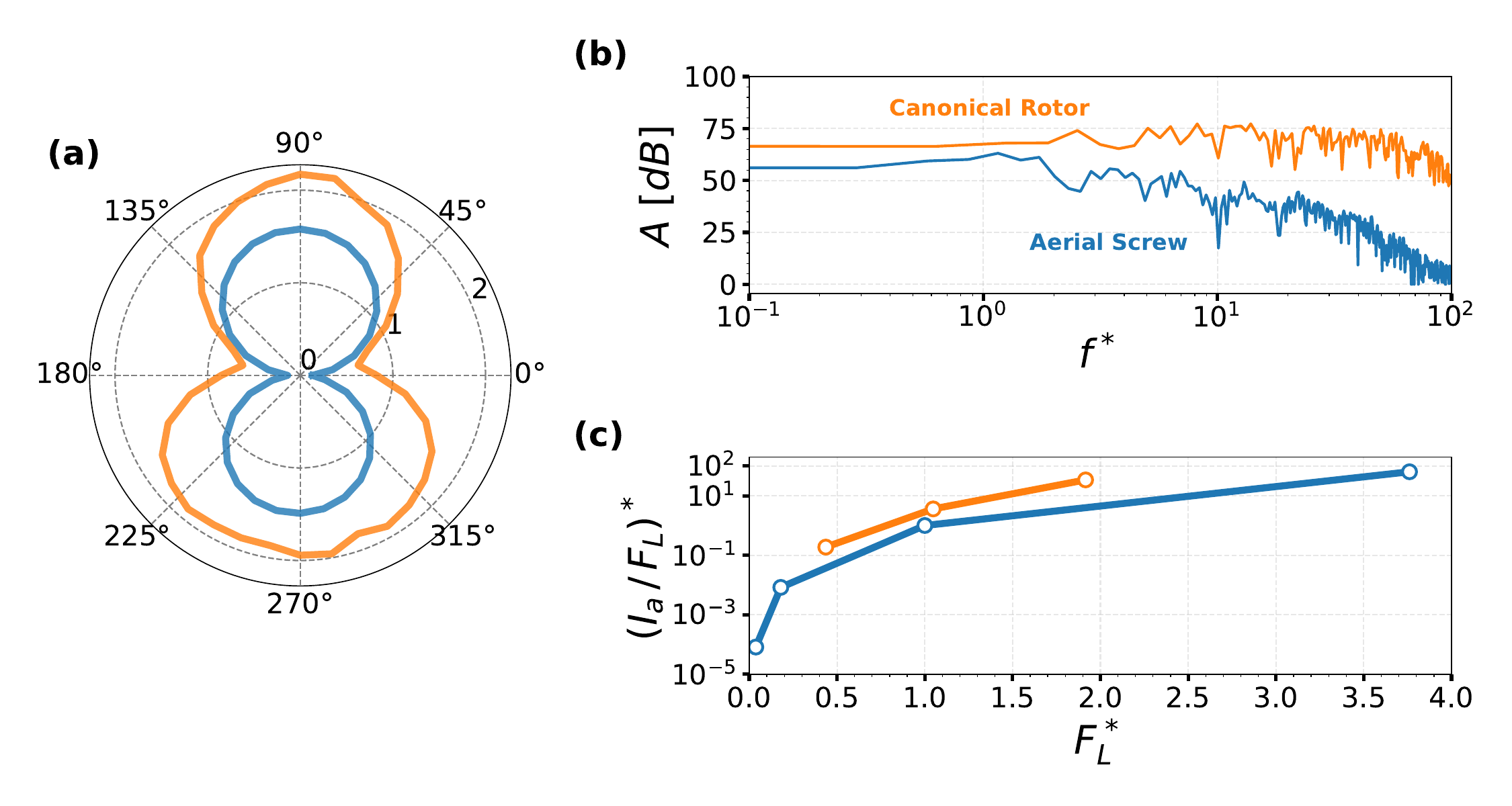}
\caption{\centering The aerial screw at Reynolds number of 8000 and the canonical rotor at Reynolds number of 12118 is compared showing (a) the directivity pattern and (b) the noise spectrum vertically above the rotation center. (c) Characteristic curve showing the acoustic intensity vs mean lift for all the aerial screw and the canonical rotor cases. The mean lift, mean directivity, and acoustic intensity of the aerial screw at Reynolds number of 8000 is used to normalize all the values.  The acoustic noise is recorded at a distance of $10R_b$ form the rotation center for all cases.}
\label{fig:comp_char_acoustics}
\end{figure*}

We now compare the aeroacoustics of both the rotors and we do this for the condition where both rotors generate similar lift. The rotors are dimensionalized using the same length and frequency scale and end up with a base radius ($R_b$) of 0.152 meter (6 inches) which is quite realistic for a small drone. The rotation speed of the aerial screw is 5000 RPM while for the canonical rotor, the rotation speed is 7550 RPM. The noise is recorded at a distance of $10R_b$ from the center of rotation (for aerial screw, the vertical location is fixed at the mid-pitch). The directivity pattern is shown in figure \ref{fig:comp_char_acoustics}a showing that the aerial screw produces less acoustic noise in all the directions. The aerial screw is a single bladed rotor design, however, due to its spiral shape, the noise is still canceled in the plane of rotation. The noise spectrum vertically above the rotation center for aerial screw at Reynolds number of 8000 and canonical rotor at Reynolds number of 12118 is shown in figure \ref{fig:comp_char_acoustics}b and we see that the screw has lower tonal and broadband noise. The broadband noise spectrum of the aerial screw also attenuates more rapidly due to the turbulent-like vortical structures being shed in the wake of the screw.  Fig. \ref{fig:comp_char_acoustics}c shows the overall acoustic intensity for all the cases and we note that sound intensity per unit lift of the aerial screw is significantly smaller than the canonical rotor at low lift values (i.e. for smaller drones) but it shows a similar trend and decreasing difference from the canonical rotor as the Reynolds number is increased. Nevertheless, this lift-specific sound intensity of the aerial screw is always smaller than that of the canonical rotor pointing to its aeroacoustic benefits.

These results are also summarized in table \ref{table:comp_between_rotors} and affirm our previous results~\citep{bioinspired_bnb} that increase in  rotor area with a concomitant decrease in rotation speed has benefits for both mechanical power and acoustic emissions. The differences in the performance of both these rotors are substantial, and in application where drone noise is a concern, unconventional rotors such as those described here and in the previous study, could be worth considering. 
\begin{table*}
\caption{\label{table:comp_between_rotors}The mechanical power and acoustic intensity for the aerial screw and the canonical two blade rotor producing similar amounts of lift are compared with star superscript indicating the quantities are normalized using the aerial screw rotor at Re 8000.}
\begin{ruledtabular}
\begin{tabular}{lccccc}
\textbf{Rotor} & \textbf{Re} &  \textbf{$\bar{C}_L$} & \textbf{$\bar{F}_L^*$}&\textbf{$(\bar{W}_M / \bar{F}_L)^*$}  &\textbf{$(I_{a} / \bar{F}_L)^*$}\\
\hline
Aerial Screw   & 2000 & 0.05 & 0.04 & 0.49 & 8.15$\times 10^{-5}$\\
Aerial Screw   & 4000 & 0.06 & 0.18 & 0.65 & 8.3$\times 10^{-3}$\\
Aerial Screw   & 8000 & 0.09 & 1 & 1 & 1\\
Aerial Screw   & 16000 & 0.08 & 3.76 & 1.87 & 64.22\\
\hline
Canonical Rotor   & 8000 & 0.37 & 0.43 & 1.12 & 0.18\\
Canonical Rotor    & 12118 & 0.38 & 1.05 & 1.73 & 3.61\\
Canonical Rotor    & 16000 & 0.40 & 1.91 & 2.36 & 34.25\\
\end{tabular}
\end{ruledtabular}
\end{table*}

\subsection{Noise Source Analysis}

In the previous section \ref{sec:aeroacoustic}, we discussed the aeroacoustic noise generated at a distance of $10R_b$ from the rotation center and we note in figure \ref{fig:comp_char_acoustics}a that the aeroacoustic noise is the largest in the vertical direction. Since the noise at a far-field location is also important in the context of drone rotors~\cite{DroneNoiseRev} and the drone rotor noise in the vertical direction are larger than noise in rotation plane and will travel directly to the ground to create annoyance, in this section we will identify and analyze the vortical structures that are primarily responsible for the far-field noise in vertical (or lift) direction. 

At low Mach numbers, the aerodynamic force on the rotor surface, $\mathbf{L}$ is the major source of noise\cite{zorumski_compact_noise} (see Eq.\ref{fwheqn}) and we use the acoustic partitioning method that was proposed by by~\citet{Seo2022} who used the Ffowcs Williams - Hawkings equation~\cite{zorumski_compact_noise} written for a compact source to calculate the far-field noise generated by the forces acting on the body ($\mathbf{F}_R=\int {\mathbf{L}} dS$) as,
\begin{equation}
    p^{'} = \frac{1}{4\pi}\left( \frac{{\bf \dot{F_R} \cdot \hat{r}}}{cr} +\frac{{\bf F_R\cdot \hat{r}}}{r^2} \right)_{t-\frac{r}{c}}\,\, ,
\end{equation}
where $r$ is the distance to the recording location, ${\bf \hat{r}}$ is the unit normal pointing from source to the recording location and $c$ is the speed of the sound. Note that when the recording location is in the lift direction, ${\bf F_R\cdot \hat{r}}=F_L$,  where $F_L$ is the lift force generated by the rotor.
In the acoustic partitioning method, the force is further decomposed by using the force partitioning method~\cite{menon2021a,menon2021b,menon2021c} (FPM) into four components, namely the force due to vortices, added mass effect, viscous diffusion of the momentum, and due to the acceleration of the body. This is done by first solving the Laplace equation for the influence field ($\phi$) given by,
\begin{equation}
     \nabla^2\phi=0 \text { in } V;  \text{   with} \, \,     \nabla\phi \cdot {\bf n}=  \begin{cases}
    n_z & \text{on $B$},\\
    0 & \text{on $\Sigma$}\, ,
  \end{cases}
\end{equation}
where, $B$ and $\Sigma$ are the body and domain boundary respectively, $V$ is the domain volume, and $n_z$ is the normal in the lift direction since we are decomposing the lift force. The gradient of this influence field is then projected onto the Navier-Stokes equation and the derivation along with the details of the FPM can be found in~\citet{menon2021a,menon2021b,menon2021c}. 
For the present rotor flows, we have found that the major force contribution comes from the force due to the vortices. For instance, if we consider the aerial screw, the vortex induced lift force accounts for 96\% of the net lift force while the force due to the viscous diffusion of the momentum accounts for only 4\% of the net lift force and the remaining two force terms are 0. The reasoning for this can be found in~\citet{prakhar2025vortices}. Thus, for the present study, the vortex induced lift force is a major source of aeroacoustic sound generation. This vortex induced lift force ($F_Q$) is defined as,
\begin{equation}
    F_Q = \int_V -2\rho \phi Q dV \,\, = \int_V f_Q dV ,
    \label{eqn:F_Q}
    \end{equation}
where, $Q$\ is the second invariant of velocity gradient defined as $\frac{1}{2} (|| \bm{\Omega}||^2- || {\bf S}||^2)$ with ${\bf S}$ and $\bm{\Omega}$ denoting the symmetric and anti-symmetric component of the velocity gradient tensor\cite{Qcrit}, and $f_Q$ is the vortex induced lift force density. Thus, the acoustic noise generated by the vortex induced force can be written as,
\begin{equation}
    p_{Q}^{'} = \frac{1}{4\pi}\left( \frac{{\bf \dot{F_Q} \cdot \hat{r}}}{cr} +\frac{{\bf F_Q\cdot \hat{r}}}{r^2} \right)_{t-\frac{r}{c}}\,\, ,
    \label{eqn:pQ}
\end{equation}
Some applications of this method can be found in \citet{Seo2022,seo2023vortex,mfpm_2025}.

Here, we employ this method to represent the aeroacoustic noise at a given far-field location into a volume source contribution where all the vortical structures are associated with the portion of the noise they generate. 
For this, we use the equation \ref{eqn:F_Q} and rewrite the vortex induced sound pressure (equation \ref{eqn:pQ}) as,
\begin{equation}
    p_{Q}^{'} = \int_V \left[ \frac{1}{4\pi} \left( \frac{{\bf \dot{f_Q} \cdot \hat{r}}}{cr} +\frac{{\bf f_Q\cdot \hat{r}}}{r^2} \right)_{t-\frac{r}{c}} \right] dV = \int_V \hat{p_Q} dV \,\, ,
\end{equation}
where $\hat{p_Q}$ describes the ``specific sound pressure'' at each point in the domain volume, and integrating this over the entire domain will yield the total sound pressure ($p_{Q}^{'}$). 
Thus, spatial distribution of $\hat{p_Q}$ indicates the major source region or the vortical structure most responsible for the noise generation.
The volume integral can also be restricted to a specific vortical structure to get the far-field noise due to a specific vortex as well. We will use this specific sound pressure to color the vortices and identify the vortical structures that are responsible for most of the noise generation.

We apply this method to both the aerial screw and the canonical rotor at Reynolds number of 8000 and 12118 respectively (i.e. when both these rotors produce similar lift force). The recording location is set at a distance of $50R_b$ in the vertical direction measured from the rotation center since we need larger recording distance to keep the point source assumption valid. 
\begin{figure*}
\centering
\includegraphics[width=\textwidth]{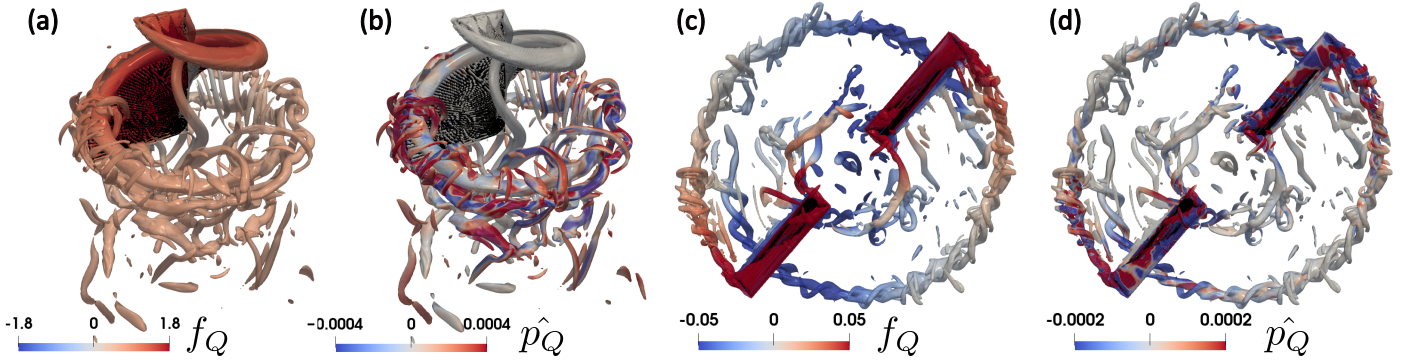}
\caption{\centering The flow field is shown using Q iso-surface and colored using (a) $f_Q$, and (b) $\hat{p_Q}$ for the aerial screw at $t/T$=5.89 . Similarly flow field for the canonical rotor colored using (c) $f_Q$, and (d) $\hat{p_Q}$ is shown at $t/T$=4.0. The integral of $\hat{p_Q}$ will give the total sound pressure recorded at at time $(t+r/c)/T$ on the recording location, where $r/c$ accounts for time delay required for propagation of sound waves.} 
\label{fig:fpQ}
\end{figure*}
The vortex induced lift force density is shown for the aerial screw and the canonical rotor in figures \ref{fig:fpQ}a and \ref{fig:fpQ}c respectively and we note that the entire spiral outer edge of the screw produces a major portion of the lift force while for the canonical rotor, this region is located along the leading edge and near the tip. The specific sound pressure for the aerial screw (figure \ref{fig:fpQ}b) reveals that while the spiral leading edge vortex generates greater amount of lift compared to smaller shed ``c'' shaped vortices near the base of the screw, this leading edge vortex provide temporally invariant lift and hence do not contribute much to the acoustic noise. The shed vortices in the wake and downwash region, while they are not large contributors to the lift force, they create larger pressure fluctuations on the rotor surface as these vortices shed and thus they are responsible for most of the acoustic noise. 
We contrast this to the specific sound pressure of the canonical rotor shown in figure \ref{fig:fpQ}d and we note that the leading edge vortex that is responsible for most of the lift force and since they also temporally vary in magnitude (this was also verified using the time derivative of the flow field), the leading edge vortex is also a significant source of the noise. The shedding at the tip and a small region in the wake of the tip also contributes to the acoustic noise since the shedding would create pressure fluctuations on the rotor surface. The shed vortical structures in the wake away from the rotor surface does not produce much lift force or the associated aeroacoustic noise since the influence field ($\phi$) decays rapidly away from the surface.

\section{Conclusions}
This study investigates the aerodynamic and aeroacoustic performance of da Vinci’s aerial screw rotor using high-fidelity direct numerical simulations. The rotor geometry was constructed based on prior design studies, and simulations were conducted across a range of Reynolds numbers to evaluate lift, mechanical power, and acoustic emissions. Our analysis confirms that while the aerial screw has a lower lift coefficient compared to a canonical two-bladed rotor, it requires significantly less mechanical power and generates substantially lower acoustic intensity for the same net lift. These advantages are primarily attributed to the screw’s large wetted area and lower rotational speed, consistent with established scaling relationships. Moreover, the continuous single-blade design of the aerial screw mitigates blade–vortex interaction noise, a key contributor to rotor aeroacoustics. We also note that the temporally invariant magnitude of the leading edge vortex of the screw helps to generate lift without significant contribution to the far-field noise. While the design is not optimized for performance, the results underscore the potential benefits of unconventional rotor geometries for noise-sensitive applications. Future work could explore geometrical variations (e.g., increasing the number of turns) and extend the analysis to higher Reynolds numbers using turbulence modeling approaches. We expect similar improvement in mechanical power and acoustic intensity for the higher Reynolds number aerial screw simulations as well since  the lift coefficient has already converged for both the aerial screw and the canonical rotor cases presented here and therefore the scaling law presented in section \ref{sec:scaling} should hold. Additionally, practical considerations such as structural integrity and dynamic stability need to be examined before translating such designs into viable rotorcraft concepts.

\section{Acknowledgments}
The authors acknowledge support from the Army Research Office
(Cooperative Agreement No. W911NF2120087) for this work. Computational resources for this work were provided by the 
high-performance computer time and resources from the DoD High Performance Computing Modernization Program.

\section{Author Declarations}
The authors have no conflicts to disclose

%

\appendix
\section{Grid Convergence}
\label{app:grid_convergence}

We use the case with the Reynolds number of 8000 to verify grid convergence where a medium mesh contains 13.8 million grid points and a fine mesh with 32 million grid points is used. The region around the rotor is placed in a uniform fine mesh with a gradually expanding mesh used away from the rotor body. In the plane of the rotation, a $2.4R_b\times 2.4R_b$ region centered around the rotor is finely resolved while in the vertical axis, a length of $2R_b$ surrounding the screw is finely resolved (figure \ref{fig:fine_mesh}).
\begin{figure}
\centering
\includegraphics[width=0.45\textwidth]{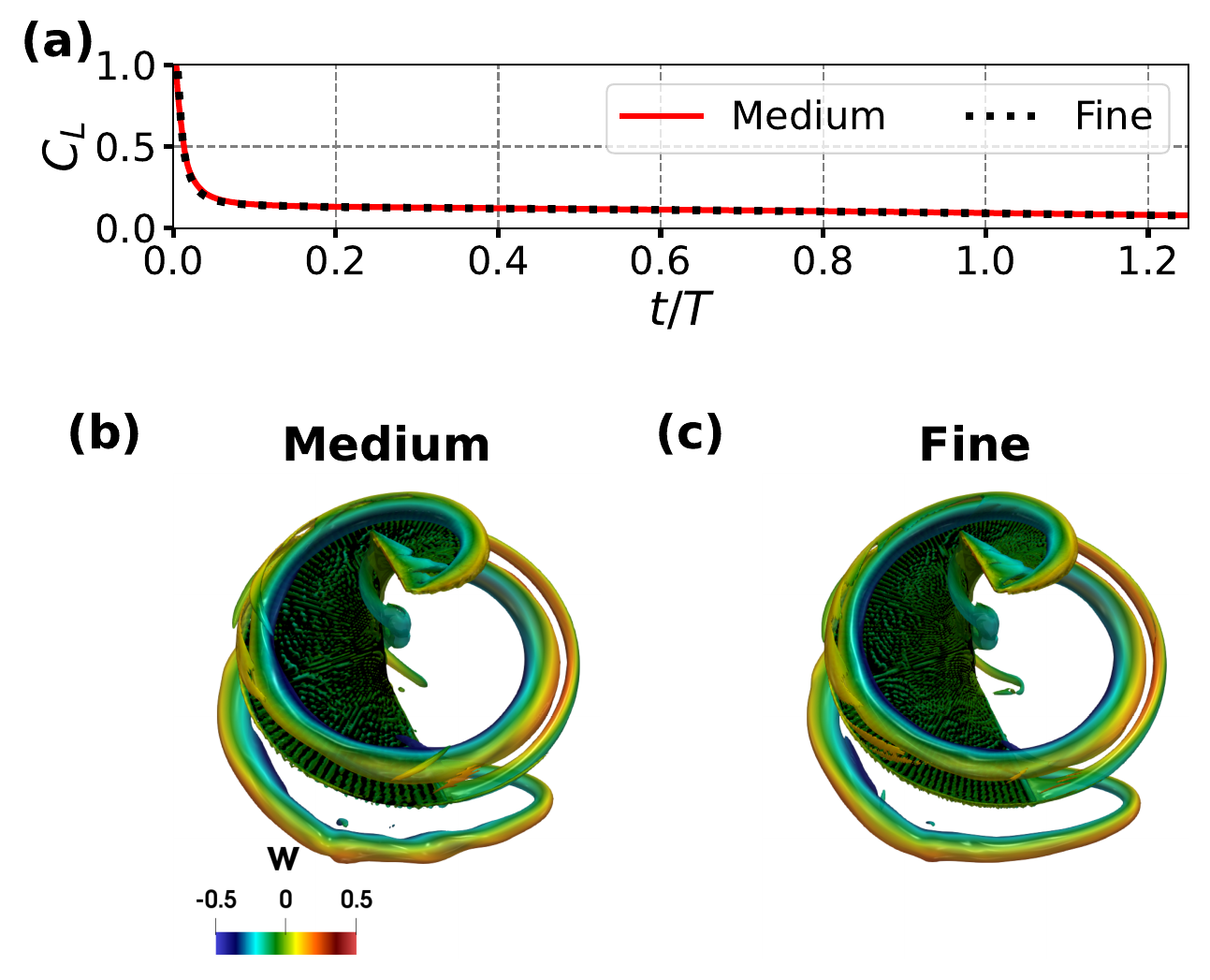}
\caption{\centering (a) The grid convergence shown using the lift coefficient corresponding to the medium and fine mesh for the case with Reynolds number of 8000. The flow field is shown using iso-surface of Q field and colored using the vertical velocity component, corresponding to the (b) medium and (c) fine mesh at $t/T=1.25$}
\label{fig:gridConvergence}
\end{figure}

The lift coefficient is shown for the case with Reynolds number of 8000 in figure \ref{fig:gridConvergence}a and the error between the RMS value of the lift coefficient between these two mesh is 1.04\%, showing that the results presented here are grid independent. The vortex field is also shown for the medium mesh (figure \ref{fig:gridConvergence}b) and the fine mesh (figure \ref{fig:gridConvergence}c) case with the vortices shown using the iso-surface of $Q$. We observe that the flow field looks visually similar on both the meshes indicating that not just the force but also the vortex structure are well converged. All the results for Reynolds 8000 and 16000 case employ fine mesh while those with Reynolds number of 4000 and 2000 employ medium size mesh.

\begin{figure}
\centering
\includegraphics[width=0.45\textwidth]{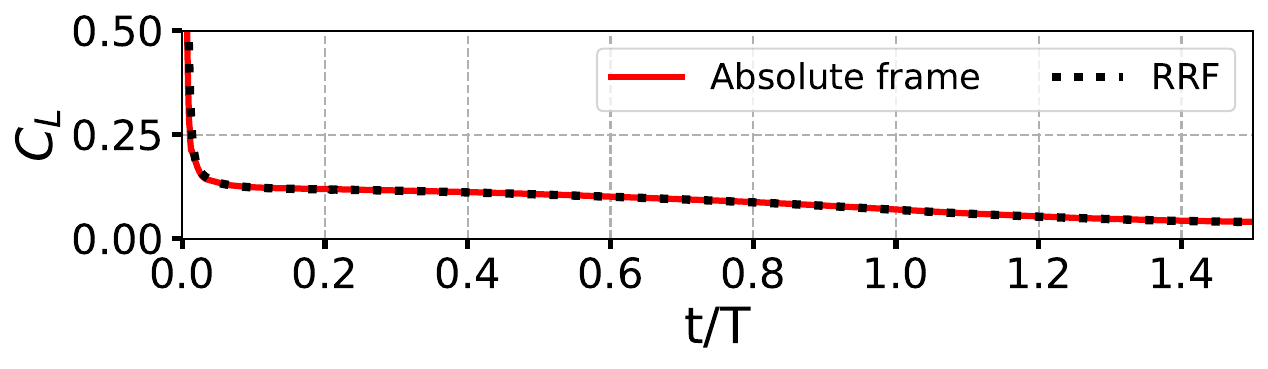}
\caption{\centering The lift coefficient is shown for the aerial screw at Reynolds number of 2000 and is simulated both in absolute frame and non-inertial rotating reference frame.} 
\label{fig:RRF_val}
\end{figure}

The rotating reference frame implementation (equation \ref{momeqn}) is validated by simulating the aerial screw at Reynolds number of 2000 case till $t/T$ = 1.25 in both absolute and non-inertial rotating reference frame and the lift coefficient is shown for both the case in figure \ref{fig:RRF_val}. 

\end{document}